\documentclass[prc,showpacs,superscriptaddress,epsf]{revtex4}

\newcommand{\be}{\begin{equation}}
\newcommand{\ee}{\end{equation}}
\usepackage{graphicx}
\usepackage{psfig}
\usepackage{epsfig}
\begin{document}
\title{ On the nuclear symmetry energy and the neutron skin in
neutron rich nuclei}
\author{A.E.L. Dieperink}
\affiliation{KVI, Zernikelaan 25, NL-9747AA Groningen, The
Netherlands}
\author{Y. Dewulf}
\author{ D. Van Neck}
\author{ M. Waroquier}
\affiliation {Laboratory of Theoretical Physics, Ghent University,
Proeftuinstraat 86, B-9000 Gent, Belgium}
\author{ V. Rodin}
\affiliation{Institut f\"ur Theoretische Physik der Universit\"at T\"ubingen
D-72076  T\"ubingen, Germany}
\date{\today}
\begin{abstract}
The symmetry energy  for nuclear matter and its relation to the
neutron skin in finite nuclei is discussed. The  symmetry energy
as a function of density obtained in a self-consistent Green
function approach is presented and compared to the results of
other recent theoretical approaches. A partial explanation of the
linear relation between the symmetry energy and the neutron skin
is proposed. The potential of several experimental methods to
extract the neutron skin is examined.
\end{abstract}
%%%%%%%%%%%%%%%%%%%%%%%%%%%%%%%%%%%%%%%
\maketitle
\section{Introduction}
The nuclear symmetry energy  plays  a central role in a variety of
nuclear phenomena. It determines to a large extent the equation of
state (EoS) and the proton fraction of neutron stars \cite{Latt},
the neutron skin in heavy nuclei \cite{Furn}, it enters as an
input in the analysis of heavy ion reactions \cite{Li,Bao}, etc.
Its value at nuclear saturation density, %$\rho_0 \approx $
  $S(\rho_0=0.17 $fm$^{-3}) \approx $ 30 MeV,
seems reasonably well established, both empirically as well as
theoretically; still different parametrizations of relativistic
mean field (RMF) models
 (which
fit observables for isospin symmetric nuclei well) lead to a
relatively wide range of predictions for the symmetry energy,
24-40 MeV. However, predictions for its density dependence show a
substantially larger variation.

Recently it has been pointed out by several authors
\cite{Brown,Furn, Bodmer} that there exists a strong correlation
between the neutron skin, $\Delta R=R_n-R_p,$ and the density
derivative of the EoS of neutron matter near saturation density.
Subsequently in a more detailed analysis in the framework of a
mean field approach Furnstahl \cite{Furn}  demonstrated that in
heavy nuclei there exists an almost linear empirical correlation
between theoretical predictions in terms of various mean field
approaches to $S(\rho)$  % (and its density dependence)  % $dS/d\rho, $
(i.e., a bulk property)  and the neutron skin, $\Delta R$ (a
property of finite nuclei).

 This observation has  contributed to a renewed interest in an accurate
determination of the neutron skin in neutron rich nuclei for several reasons. %
%e.g. $^{208}$Pb,
% One aim is to constrain parameters that enter in calculations of
% the symmetry energy, \cite{Horo01}.
 First precise experimental information on the neutron skin could help
%A second issue is to try to
to further constrain interaction parameters that play a role in the
calculation of
 the symmetry energy \cite{Horo01}. Furthermore a precise value of the
neutron skin
is required as an input in several processes of physical interest, e.g.
the analysis of energy shifts in deeply bound pionic
atoms \cite{Kolo}, and in the analysis of atomic parity violation experiments
(weak charge) \cite{Poll99}.
 It has been shown that the calculated symmetry energy
is quite insensitive to details of modern realistic NN interactions \cite
{Engvik}.
% investigate the sensitivity of microscopic
%results of microscopic calculations of the symmetry energy
 On the other hand the symmetry energy and in particular its density dependence
can vary substantially with the  many-body approximations
employed. For instance the results of lowest order BHF and
variational calculations do not seem to agree well.

The aim of this  paper is threefold: First we address the
sensitivity of the symmetry energy to various many-body
approximations. To this end we present results of a calculation of
$S(\rho)$ using the self-consistent Green function (SCGF) approach
and compare the result with various other theoretical approaches.
Secondly  we will try to provide some new insight in the origin of
the ``Furnstahl" relation; for this purpose we use  the
Landau-Migdal effective
 interaction in the mean field approximation.
 Finally  in view of the large variety of existing and proposed experimental
methods to  determine the neutron skin $\Delta R$
we examine the merits of some recently proposed methods
that seem to be of potential interest to provide more accurate
information on the neutron skin  in the near future.

Section \ref{Overview} is devoted to an overview of theoretical approaches to
the symmetry energy, and a new calculation in terms of the
self-consistent Green function approach is presented. In Section
\ref{Relationship} an interpretation of the Furnstahl relation is
presented in terms of the Landau-Migdal approach. Section
\ref{Exp} contains an overview of various experimental methods to
deduce information on the neutron skin and Section
\ref{Discussion} contains a short discussion of implications for
other physical processes where the information on the neutron skin
is required as an input.
%%%%%%%%%%%%%%%%%%%%%%%%%%%%
\section{The symmetry energy in nuclear matter}
\label{Overview}
 The symmetry energy $S(\rho)$ is defined in terms of a
Taylor series  expansion of the energy per particle for nuclear
matter in terms of the asymmetry
$\alpha=(N-Z)/A $ (or equivalently the proton fraction $x=Z/A$),
%$x=\rho_p/(\rho_p+\rho_n)= (1-\alpha)/2$)
 \be E(\rho,\alpha)=
E(\rho,0) +S(\rho)\alpha^2 + O(\alpha^4)+ \ldots . \label{BE} \ee
It has been shown
\cite{Zuo,Zuo02} that deviations from the parabolic law in
Eq.(\ref{BE}), i.e. terms in $\alpha^4,$ are quite small.

Near the saturation density $\rho_0$ the energy of isospin-symmetric
matter $E(\rho,0)$  and the symmetry energy can be expanded as
 \be  E(\rho,0)= E_0+ \frac
{K}{18\rho_0^2} (\rho-\rho_0)^2 +\ldots , \ee
and
 \be S(\rho)= \frac{1}{2}
\frac{\partial^2E(\rho,\alpha)}{\partial \alpha^2}|_{\alpha=0} =
a_4+ \frac{p_0}{\rho_0^2} (\rho-\rho_0) + \frac{\Delta
K}{18\rho_0^2}(\rho-\rho_0)^2+\ldots  \label{esymm}. \ee
The parameter $a_4$ is the symmetry energy at equilibrium and the slope
parameter $p_0$ governs the density dependence.

The relevance of the nuclear matter results in part depends on the question
whether there is a surface  contribution to the symmetry energy for finite
nuclei. In ref.\cite{Oya} it  was found that the latter is of minor importance,
which has  also been confirmed in \cite{Furn}.

\subsection{Self-consistent Green function and  Brueckner approach}

In this section we describe the calculation of the symmetry energy
in the SCGF approach. Since the latter can be considered as a
generalization of the lowest order Brueckner-Hartree-Fock (BHF)
method we start with a brief discussion of the symmetry energy in
the latter case.

\subsubsection{Symmetry energy in BHF}
In the Brueckner-Hartree-Fock  approximation, the
Brueckner-Bethe-Goldstone (BBG) hole-line expansion is truncated
at the two hole-line level. The short-range NN repulsion is
treated by a resummation of the particle-particle ladder diagrams
into an effective interaction or $G$-matrix. Self-consistency is
required at the level of the BHF single-particle spectrum
$\epsilon^{BHF}(k)$,
\begin{equation}
\epsilon^{BHF}(k)=\frac{k^2}{2m}+\sum_{k'<k_F}\mbox{Re}
<kk'|G(\omega=\epsilon^{BHF}(k)+\epsilon^{BHF}(k'))|kk'> .
\label{eq:spectrum}
\end{equation}
In the standard choice BHF the self-consistency requirement
(\ref{eq:spectrum}) is restricted to hole states ($k<k_F$) only,
while the free spectrum is kept for particle states $k>k_F$. The resulting
gap in the s.p.\ spectrum at $k=k_F$ is avoided in the continuous-choice BHF
(ccBHF), where Eq.~(\ref{eq:spectrum}) is used for both hole and particle
states. The continuous choice for the s.p.\ spectrum is
closer in spirit to many-body Green's function perturbation theory.
Moreover, recent results indicate \cite{Bal01,Bal00} that the
contribution of higher-order terms in the hole-line expansion is
considerably smaller if the continuous choice is used.

The BHF energy per nucleon can be easily evaluated for both
symmetric nuclear matter (SNM) and pure neutron matter (PNM) using
the energy sum-rule,
\begin{equation}
\frac{E}{A} = \frac{d}{\rho} \int \frac{d^3k}{(2\pi)^3} \left(
\frac{k^2}{2m} + \epsilon^{BHF}(k) \right) \theta(k_F-k) ,
\label{eq:epp}
\end{equation}
where $\rho = \frac{dk_F^3}{3\pi^2}$ is the density and $d$ is the
isospin degeneracy ($d=1 (2)$ for PNM (SNM)). The symmetry energy
$S(\rho)$ is obtained as the difference between PNM and SNM
energies for the same density.

\begin{figure}[h] \centering{
\includegraphics[width=0.4\textwidth]{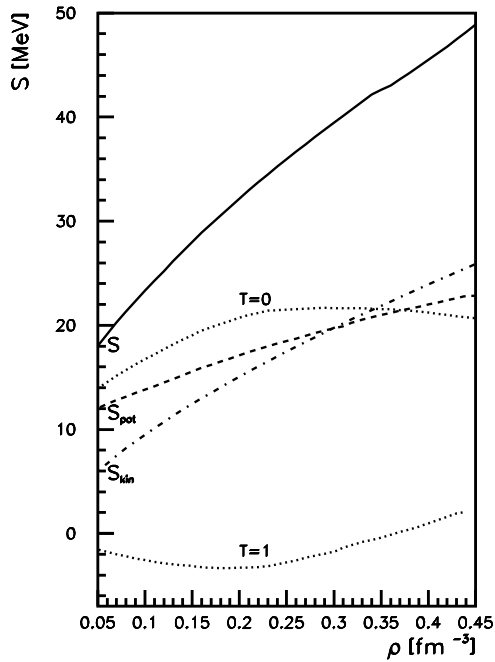}
\caption{The symmetry energy $S$ (full line) and the contributions
to $S$ from the kinetic (dash-dotted line) and potential energy
(dashed line), calculated  within a ccBHF scheme and using the
Reid93 interaction. Also shown (dotted lines) are the $T=0$ and
$T=1$ components of the potential energy contribution.}
\label{fig:bhf}}
\end{figure}

We first performed ccBHF calculations with the Reid93
interaction, including partial waves with $J<4$ in
the calculation of the G-matrix. The results are presented in
Fig.\ref{fig:bhf}, where the symmetry energy $S$ is
decomposed into various contributions as suggested in ref.\cite{Zuo} and
shown as a function of nucleon density $\rho$.
The contribution $S_{\rm kin}$ of the kinetic energy to the BHF
symmetry energy is given by the free Fermi-gas expression
\footnote{This expression differs from the standard one, which is
based upon
 the derivative rather than  the finite difference.}
\begin{equation}
S_{\rm kin}=E_{kin,PNM}-E_{kin,SNM}=
\frac{3}{10m}(3\pi^2)^{\frac{2}{3}}\rho^{\frac{2}{3}}
\left(1-\frac{1}{2^{\frac{2}{3}}}\right),
\end{equation}
and it determines to a large extent the density dependence of $S$.
In Fig.\ref{fig:bhf} we also show the symmetry potential
$S_{\rm pot} = S -S_{\rm kin}$, which is much flatter, and the
contributions to $S_{\rm pot}$ from both the
isoscalar ($T=0$) and isovector ($T=1$) components of the
interaction. Over the considered density range $S_{\rm pot}$ is
dominated by the positive $T=0$ part. The $T=0$ partial waves,
containing the tensor force in the  $^3S_1$-$^3D_1$ channel which
gives a major contribution to the potential energy in SNM, do not
contribute to the PNM energy. The $T=0$ contribution peaks at
$\rho\approx 0.3$ fm$^{-3}$. The decrease of this contribution at
higher densities is compensated by the increase of the $T=1$
potential energy, with as a net result a much weaker
density-dependence of the total potential energy.
(It should be
noted that inclusion of a three-nucleon interaction in general
leads to a substantial increase for the slope parameter $p_0$
\cite{Zuo02}.)

\subsubsection{Symmetry energy in SCGF approach}
In recent years several groups have considered the replacement of
the BBG hole-line expansion with self-consistent Green's function
(SCGF) theory \cite{Ramos,Bozek,Dick, Dewulf01,Dewulf02, Dewulf03}. In
\cite{Dewulf02,Dewulf03} the binding energy for SNM was calculated
within the SCGF framework and using the Reid93 potential. In the
present paper we have extended these calculations to PNM and
calculated the corresponding symmetry energy. Details of a
technical nature can be found in \cite{Dewulf02}.

A SCGF calculation differs in two important ways from a BHF-
calculation. Firstly, within SCGF particles and holes are treated
on an equal footing, whereas in BHF only intermediate particle
$(k>k_F)$ states are included in the ladder diagrams. This aspect
ensures thermodynamic consistency, e.g.\ the Fermi energy or
chemical potential of the nucleons  equals the binding energy at
saturation (i.e., it fulfills the Hugenholz-van Hove theorem). In
the low-density limit BHF and SCGF coincide. As the density
increases the phase space for hole-hole propagation is no longer
negligible, and this leads to an important repulsive effect on the
total energy. Secondly, the SCGF generates realistic spectral
functions, which are used to evaluate the effective interaction
and the corresponding nucleon self-energy. The spectral functions
include a depletion of the quasi-particle peak and the appearance
of single-particle strength at large values of energy and
momentum, in agreement with experimental information from
$(e,e'p)$ reactions. This is in contrast with the BHF approach
where all s.p.\ strength is concentrated at the BHF-energy as
determined from Eq. (\ref{eq:spectrum}).

In a SCGF approach the particle states ($k>k_F$), which are absent
in the BHF energy sum rule of Eq.~(\ref{eq:epp}), do contribute
according to the energy sum rule
\begin{equation}
\frac{E}{A} = \frac{d}{\rho} \int \frac{d^3k}{(2\pi)^3}
\int_{-\infty}^{\varepsilon_F} d\omega\ \left( \frac{k^2}{2m} +
\omega \right) S_h(k,\omega) , \label{eq:epps}
\end{equation}
expressed in terms of the nucleon spectral function $S_h
(k,\omega)$.

\begin{figure}
\centering{
\includegraphics[width=0.8\textwidth,height=0.5\textwidth] {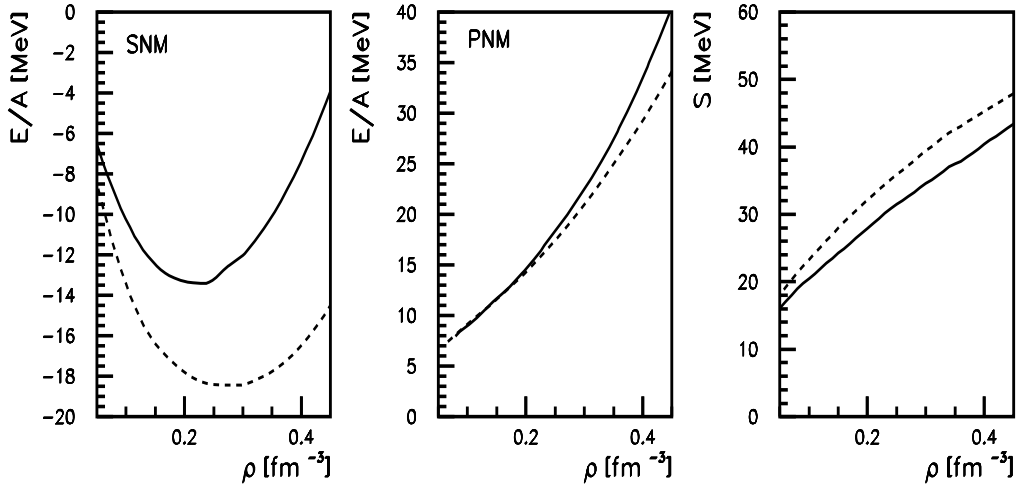}
\caption{The total-energy per particle for symmetric nuclear matter (left
panel) and pure neutron matter (central panel) for the Reid93-interaction.
The dashed line refers to a ccBHF calculation, the full line to a SCGF
calculation. The right panel displays the symmetry energy in these two
approaches.}
\label{fig:symbhf}}
\end{figure}

The results for the ccBHF and SCGF calculations for both SNM and
PNM are compared in the left and central panels of
Fig.\ref{fig:symbhf} for the Reid93 interaction. The
inclusion of high-momentum nucleons leads roughly to a doubling of
the kinetic and potential energy in SNM, as compared to BHF.
As seen in Fig.\ref{fig:symbhf}, the
net result for the total energy of SNM is a repulsive effect,
increasing with density \cite{Dewulf02}. This leads to a stiffer
equation of state, and a shift of the SNM saturation density
towards lower densities. The above effects are dominated by the
tensor force (the isoscalar $^3S_1-^3D_1$ partial wave).
Consequently, the effects are  much smaller in PNM.

The corresponding symmetry energy, shown in the right panel of
Fig.\ref{fig:symbhf}, is dominated by the shift in the total
energy for SNM, and lies below the ccBHF symmetry energy
in the entire density-range. At $\rho_0 =0.16$fm$^{-3}$ the
symmetry energy parameter $a_4$ is reduced from
28.9 MeV to 24.9 MeV, while the slope $p_0$ remains almost the
same (from 2.11 MeVfm$^{-3}$  to 1.99 MeVfm$^{-3}$).

\subsection{Comparison of symmetry energy in  other approaches }

\subsubsection{Calculations with realistic NN forces}

Engvik et al.\ \cite{Engvik} have performed lowest-order
Brueckner-Hartree-Fock calculations in
SNM and PNM for all ``modern'' potentials (CD-Bonn, Argonne v18,
Reid93, Nijmegen I and II), which fit the Nijmegen NN scattering
database with high accuracy. They concluded that for small and
normal densities the symmetry energy is largely independent of the
interaction used, e.g.\ at $\rho_0$ the values of $S$ vary around
an average value of $a_4$=29.83 MeV by about 1 MeV. At larger
densities the spread becomes larger; however, the symmetry energy
keeps increasing with density, in contrast to some of the older
potentials like Argonne v14 and the original Reid interaction
(Reid68) for which $S(\rho )$ tended to saturate at densities
larger than $\rho =0.4$ fm$^{-3}$.
Some insight into the microscopic origin of the symmetry potential
was obtained by Zuo et al. \cite{Zuo} who decomposed the symmetry
energy into  contributions from kinetic and potential energy. The
BHF calculations in \cite{Zuo} used the Argonne v14 and the
separable Paris interaction. In Fig.\ref{fig:bhf} we showed that the
use of the modern Reid93 potential leads to essentially the same conclusions.

%Results with the self-consistent Green function approach differ
%from BHF by a reduced value of $a_4$, whereas the slope $p_0$
%remains unchanged (see Fig.\ref{fig:symbhf}).

Detailed studies for SNM and PNM using variational chain summation
techniques were performed by Wiringa et al.\ \cite{Wir} for the
Argonne Av14 and Urbana Uv14 NN interaction, in combination with
the Urbana UVIII three-nucleon interaction (TNI), and by Akmal et
al.\ \cite{Akmal} for the modern Av18 NN potential in combination
with the UIX-TNI. Results for $a_4$ and $p_0$, extracted from
\cite{Wir} and \cite{Akmal}, are shown in Table~\ref{variational}.
The inclusion of TNI stiffens the EoS for both SNM and PNM, and
increases considerably the value of the symmetry energy $a_4$ and
its slope $p_0$ at the empirical density $\rho_0$. The effect of
including a relativistic boost correction $\delta v$ (in
combination with a refitted TNI) on the values of $a_4$ and $p_0$
is sizeable as well.

\begin{table}[table:ebinding]
\caption{Results for the symmetry energy parameters $a_4$ and
$p_0$ from the variational calculations of
Refs.\cite{Wir},\cite{Akmal} using the Argonne and Urbana NN
potentials, in combination with Urbana models for the
three-nucleon interaction. The last column includes a relativistic
boost correction $\delta v$ and the adjusted UIX* three-nucleon
interaction.}
\begin{center}
\begin{tabular}{c|c|c|c|c|c|c|c}
&Av14 &Av14+UVIII & Uv14 & Uv14+UVIII & Av18 &
Av18+UIX & Av18+$\delta v$+UIX* \\
\hline
$a_4$ [MeV]&24.90&27.49&26.39&28.76&26.92&29.23&30.1\\
$p_0$ [MeV fm$^{-3}$]&2.02&2.71&2.38&3.04&1.95&3.24&2.95\\
\end{tabular}
\end{center}
\label{variational}
\end{table}

The symmetry energy has also been computed in the
Dirac-Brueckner-Hartree-Fock (DBHF) approach \cite{Lee, Toki}. In
relativistic  approaches the symmetry energy generally is found to
increase almost linearly with density, and more rapidly than in
the non-relativistic case. This difference can be attributed to
two effects. First the covariant kinetic energy which is inversely
proportional to $\sqrt{k_F^2+ m^{*2}}$  is larger because of the
decreasing Dirac mass, $m^*$, with increasing density. Secondly
the contribution from rho-exchange appears to be larger than in
the non-relativistic case \cite{Lee}.

\subsubsection{Mean-field approach using effective interactions}

Since the Furnstahl relation has been verified mainly in terms of
mean-field models we discuss some results obtained in these
approaches, which in general are based upon a parameterized
effective interaction.

Brown \cite{Brown} has investigated proton and neutron radii
in terms of the non-relativistic Skyrme Hartree-Fock (SHF) model.
First he noted that a certain combination of parameters in the SHF
is not determined well by a fit to ground state binding energies,
and that  a wide range of predictions for the  EoS for PNM is
obtained. He also pointed out a direct correlation between the
derivative of the neutron matter EoS (i.e., basically the symmetry energy
coefficient $p_0$) and the neutron skin in $^{208}$Pb.

Covariant approaches are in general based upon either a covariant
lagrangian with $\sigma,\ \omega$ and $\rho$ exchange (and
possibly other mesons) \cite{Ring, Horo02}, or on the use of
contact interactions \cite{Furn}, solved as an energy density
functional (EDF) in the Hartree-Fock approximation. Sets of  model
parameters are determined by fitting bound state properties of
nuclei. Specifically the isovector degree of freedom is determined
by the exchange of isovector mesons; in case of rho-meson exchange
 the (positive definite)  contribution to $S$ is given by
 \be a_4= \frac{k^2_F}{6 \sqrt{m^{*2}+k_F^2}}+ \frac{g^2_\rho}{8m^{2}_\rho}
\rho_0,
 \ee
and its potential energy contribution to $p_0$, which scales with
that for $a_4$, is
   $ \frac{g^2_\rho}{8m^{2}_\rho}$  \cite{Furn}.
Typical values obtained for $p_0$ are around 4-6 MeV fm$^{-3}$,
and $ a_4\sim $ 30-36 MeV, i.e. considerably larger than in
non-relativistic approaches
 (a large part of the enhancement can
be ascribed to the fact that the kinetic contribution is larger,
because $m^* <m$).  Recently in \cite{Liu02,Greco} this approach was
extended  by inclusion of  the isovector-scalar partner, $\delta,$
of the isoscalar scalar $\sigma-$meson. Because of the presence of
the Lorentz factor $m^*/E$ in the scalar potential contribution,
$\sim -\frac{g_\delta^2}{8m_\delta^2} \frac{m^*}{E}, $ which
decreases with increasing density its inclusion leads to an even
larger net value for $p_0$ \cite{Liu02,Greco}.

 Since  explicit pion exchange is usually not included in the mean
field approaches it is difficult to make a meaningful comparison
with microscopic  ones. In fact it can be argued that in contrast
to isoscalar properties the long-range pion exchange could play an
essential role in determining the isovector properties
\cite{Furn}.

%%%%%%%%%%%%%%%%%%%%%%%%%%%%%%%%%%%%%%%%%%%%%%%%%%
\subsubsection{Effective field theory}
Recently the density dependence of the symmetry energy has been
computed in chiral perturbation effective field theory, described
by   pions plus one cutoff parameter, $\Lambda,$ to simulate the
short distance behavior \cite{Kais}. The nuclear matter
calculations have been performed up to three-loop order; the
resulting EoS is expressed as an expansion in powers of $k_F,$ and
the value of $\Lambda \approx 0.65$ GeV is adjusted to the
empirical binding energy per nucleon. The value obtained in this
approach for $a_4= 33$ MeV is in remarkable agreement with the
empirical one; %  and $p_0=..$ ;
at higher densities ($\rho > 0.2$ fm$^{-3}$)  a downward bending is predicted.
However, in its present form the validity of this approach is clearly confined
to relatively small values of the Fermi momentum, i.e. rather low densities.
 It is interesting to note that there
are relatively small (large) contributions to $a_4$ coming from
one-pion exchange Fock diagram (three-loop diagrams with either two
or three medium insertions).

\subsubsection{Comparison}
\begin{figure}
\centering{
\includegraphics[width=0.5\textwidth,height=0.7\textwidth] {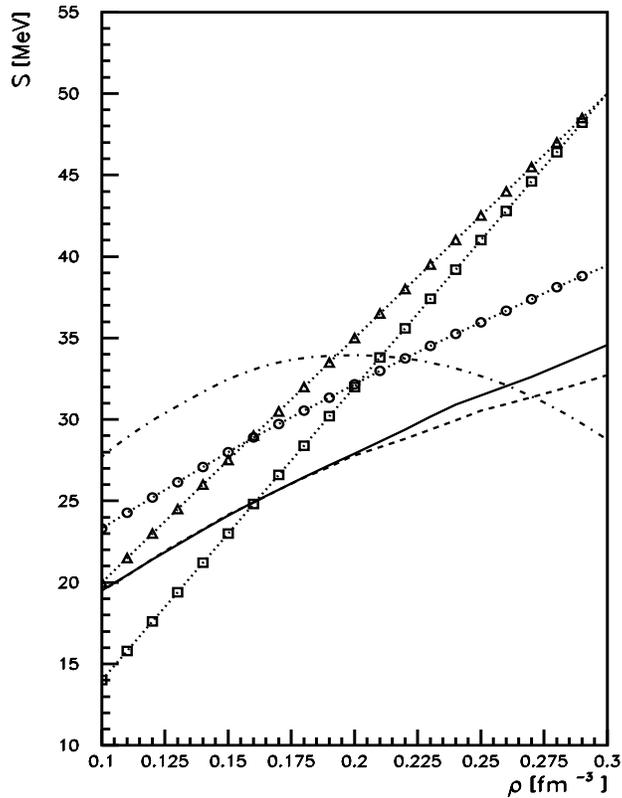}
 \label{Overview-Srho}
%\caption{
\caption{ Overview of several theoretical predictions for the
symmetry energy $S$: Brueckner-Hartree-Fock (continuous choice)
with Reid93 potential (circles), self-consistent Green function
theory with Reid93 potential (full line), variational calculation
from \cite{Wir} with Argonne Av14 potential (dashed line),
Dirac-Brueckner-Hartree-Fock calculation from \cite{Lee}
(triangles), relativistic mean-field model from \cite{Liu02}
(squares), effective field theory from \cite{Kais} (dash-dotted
line).} }
\end{figure}

To summarize the present status in Fig.(\ref{Overview-Srho})
various results of the approaches for $S(\rho)$ discussed above
are compared. As noted above one sees that the covariant models
yield a much larger increase of $S$ with the density than the
non-relativistic approaches. The LOBHF leads to a higher value of
$S$ than both variational and the SCGF method which include more
correlations; that the SCGF result is close to the variational
approach may be fortuitous. Effects from three-body forces are not
included.

%%%%%%%%%%%%%%%%%%%%%%%%%%%%%%
\section{Relationship between symmetry energy and $\Delta R$}
\label{Relationship}
 Brown \cite{Brown} and Furnstahl \cite{Furn} have pointed out that within the
framework
 of mean field models there exists an
almost linear  empirical correlation between theoretical
predictions for both $a_4$ and its density dependence, $p_0, $    and
 the neutron skin, $\Delta R=R_n-R_p,$ in heavy nuclei.
% This is illustrated for $^{208}$Pb in Fig. 3 (from \cite{Furn}).
 This observation suggests an intriguing
relationship between a bulk property of infinite nuclear matter
and a surface property of finite systems.
\\ Here we wish to address this question from a different point of view,
namely in the spirit of Landau-Migdal approach.
Let us consider a simple mean-field model (see, e.g., \cite{Gor00})
with the Hamiltonian consisting of the single-particle mean field
part $\hat H_0$ and the residual particle-hole interaction $\hat
H_{p-h}$:
\begin{eqnarray}
&\hat H=\hat H_0+\hat H_{p-h},\ \hat H_0=\sum\limits_a (T_a+U(x_a)),\\
&U(x)=U_0(x)+U_1(x)+U_C(x),\label{1.1}\\
&U_0(x)=U_0(r)+U_{so}(x);\ U_1(x)=\frac12 S_{\rm pot}(r)\tau^{(3)};\
U_C(x)=\frac12 U_C(r)(1-\tau^{(3)}), x=(r,\sigma,\tau) \nonumber
\end{eqnarray}
Here, the mean field potential $U(x)$ includes the
phenomenological isoscalar part $U_0(x)$ along with the isovector $U_1(x)$
and the Coulomb $U_C(x)$ parts calculated consistently in
the Hartree approximation; $U_0(r)$ and $U_{so}(x)=U_{so}(r)\vec{\sigma}
\cdot\vec l$
are the central and spin-orbit parts of the isoscalar mean field,
respectively; $S_{\rm pot}(r)$ is the symmetry potential
 (the potential part of the symmetry energy).
\\ In the Landau-Migdal approach the effective isovector particle-hole
interaction $\hat H_{p-h}$
is given by
\begin{equation}
\hat H_{ph}=\sum\limits_{a>b}
(F'+G'\vec\sigma_a\vec\sigma_b)\vec\tau_a\vec\tau_b
\delta(\vec{r}_a-\vec{r}_b),
\label{1.3}
\end{equation}
where $F'$ and $G'$ are the phenomenological Landau-Migdal
%isovector
parameters.

The model Hamiltonian $\hat H$ in Eq.(\ref{1.1}) preserves isospin
symmetry if the condition
 \be
%\begin{eqnarray}
\label{1.4}
[\hat H, \hat   T^{(-)}]=\hat U^{(-)}_C,  \label{commHT} \ee
is fulfilled, where
$ \hat T^{(-)}=\sum\limits_a \tau_a^{(-)},\ \hat U^{(-)}_C=\sum\limits_a
U_C(r_a)\tau_a^{(-)}. $
  With the use of  Eqs.~(\ref{1.1}),(\ref{1.3}),(\ref{1.4})
    the condition eq.(\ref{commHT})  in the RPA formalism
  leads to
a selfconsistency relation between  the   symmetry potential and
the Landau parameter $F'$ \cite{Bir74}:
\begin{eqnarray}
&S_{\rm pot}(r)=2F'n^{(-)}(r), \label{1.5}
\end{eqnarray}
where $n^{(-)}(r)=n^{n}(r)-n^{p}(r)$ is the neutron excess density.
Thus, in this model the depth of the symmetry potential is
controlled by the Landau-Migdal parameter $F'$ ( analogously
to the role played by the parameter $g_\rho^2$ in relativistic mean field
models).

\begin{figure}
\includegraphics[width=10cm]{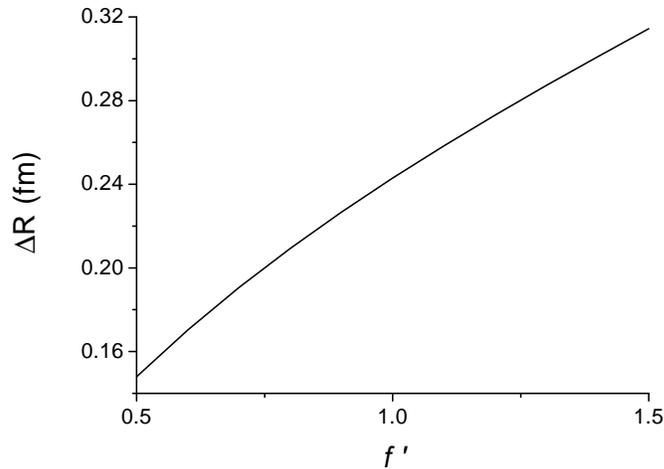}
\caption{Neutron skin in $^{208}$Pb vs. the Landau-Migdal parameter $f'$.}
 \label{DRvsF'}
\end{figure}

  $S_{\rm pot}(r)$ is obtained from Eq.(\ref{1.5}) by an iterative
procedure;
%Solving eq.(\ref{1.5})  by iteration yields $S_{\rm pot}(r),
  the resulting  dependence of $\Delta R$ on the dimensionless
parameter $f'=F'/ (300$ MeV\,fm$^3$)  shown in fig.~\ref{DRvsF'}
  indeed illustrates that
$\Delta R$ depends almost linearly on $f'.$
  Then with the use of the Migdal relation \cite{Migd} which relates symmetry
energy
  and $f',$
   \be a_4= \frac{\epsilon_F}{3}(1+2f'), \ee
 a similar almost linear correlation
between   the symmetry energy, $a_4,$ and the neutron skin is obtained.

 To get more insight in the role of $f'$ % in the "Furnstahl relation"
 we consider small variations $\delta F'$.
  Neglecting  the variation
of $n^{(-)}(r)$ with respect to $\delta F'$ one has a linear
variation of the symmetry potential: $\delta S_{\rm pot}(r)=2\delta
F'n^{(-)}(r)$. Then in first order perturbation theory, such a
variation of $S_{\rm pot}$ causes the following variation of the
ground-state wave function
\begin{equation}
|\delta 0\rangle=\delta F'\sum\limits_{s} \frac{\langle s | \hat
N^{(-)} | 0\rangle}{E_0-E_s} |s\rangle, \label{1.33}
\end{equation}
with $s$ labeling the eigenstates of the nuclear Hamiltonian and
a single-particle operator $\hat N^{(-)}=\sum\limits_a n^{(-)}(r_a)\tau^{(3)}
_a$.
Consequently the variation of the expectation value
of the single-particle operator $\hat V^{(-)}=\sum\limits_a r^2_a\tau^{(3)}_a$
with $\langle 0 | \hat V^{(-)} | 0 \rangle=N R^2_n-Z R^2_p$
 can be written as
\begin{equation}
%\delta (\langle r^2\rangle_n-\langle r^2\rangle_p)
R_p \delta( \Delta R) =\delta F'\frac{2}{A}\sum\limits_{s}
\frac{{\rm Re}\langle 0 | \hat N^{(-)} | s\rangle \langle s | \hat
V^{(-)} | 0\rangle} {E_0-E_s}. \label{deltar2np}
\end{equation}
In practice the sum in Eq.~(\ref{deltar2np}) is exhausted mainly
by the isovector monopole resonance of which the high excitation energy
(about 24 MeV in $^{208}$Pb) justifies the perturbative
consideration. We checked that Eq.~(\ref{deltar2np}) is able to
reproduce directly calculated $\delta (\Delta R)$ shown in
fig.~\ref{DRvsF'} with the accuracy of about 10\%. As a result  a
simple microscopic interpretation of the linear correlation
between the neutron skin thickness and Landau parameter $F'$ is
obtained.

%%%%%%%%%%%%%%%%%%%%%%%%%%%%%%%%%%%%%%%%%%%%%%%%%%%%%%%%%%%%%%%%
\section{Experimental methods to determine $\Delta R$} \label{Exp}
A  variety of experimental approaches have been explored in the
past to obtain information on $\Delta R.$ To a certain extent all
analysis contain a certain model dependence, which is difficult to
estimate quantitatively. It is not our intention to present a full
overview of existing methods for the special case of $^{208}$Pb.
In particular the results obtained in the past from the analysis of elastic scattering
of protons and
neutrons have varied depending upon specifics of the analysis
employed. At present the most accurate value comes from a recent detailed
analysis of the elastic proton scattering reaction at $E=0.5-1$
GeV \cite{Clark03}, and of neutron and proton scattering at
$E=40-100$ MeV \cite{Karat02}. For details we refer to these
papers. Here we restrict ourselves to a discussion of the
 some less well known methods that have the potential to provide more accurate
 information on the neutron skin in the future. In particular the
 excitation of isovector giant resonances through the restoring
 force contain information about the symmetry energy.
%%%%%%%%%%%%%%%%%%%%%%%%%%%%%%%%%
\subsection{Isovector giant resonances}
We begin with a brief overview of the study of excitation of
isovector giant resonances. Sum rules for the latter
 contain direct information on $\Delta R$.
%%%%%%%%%%
\subsubsection{Giant Dipole Resonance (GDR)}
In the past the excitation of the isovector giant dipole resonance (GDR) with
isoscalar
probes has been used to   extract $\Delta R/R$ \cite{Kras94}.
In the DWBA optical model analysis of the cross section the neutron and
proton transition densities are needed as an input.
   In the Goldhaber-Teller picture \be
g_i(r)= -\kappa \frac{2N_i}{A} \frac{d\rho_i}{dr} \ee with $\kappa
$ the oscillation amplitude and $(i=p,n)$ . Assuming ground state neutron and
proton
distributions of the form $(x=(N-Z)/A)$
\be \rho_i(r)=\frac{1}{2}(1\pm x \mp
\gamma x) \rho( r- c(1 \pm \gamma x/3)). \ee
While for $N=Z$ the
 transition density vanishes,  for $N>Z$ the
isovector transition density is finite $$ \Delta g(r) = \kappa \gamma
\frac{N-Z}{A}( \frac{d\rho}{dr}+\frac{c}{3} \frac{d^2\rho}{dr^2}),
$$ where $\gamma$ is related to $\Delta R,$ $\gamma = \frac{ 3A}{
2(N-Z) } \Delta R/ R_0. $
\\ Excitation of the GDR by alpha
particle scattering (isoscalar probe) the corresponding transition
optical potential is given by \be \Delta
U_{tr}=\kappa\gamma\frac{N-Z}{A} ( \frac{dU}{dr}+ \frac{R_0
d^2U_0}{3dr^2}). \ee By comparing the experimental   cross section
with the theoretical one  (calculated as a function of the ratio
$\Delta R/ R$) the value of $\Delta R/ R$ can be deduced.

It is difficult to make a quantitative estimate of the uncertainty
in the result coming from the model dependence of the approach. In
the analysis several assumptions must be made, such as the radial
shape of the density oscillations and about the
 actual values of the  optical model parameters.

\subsubsection{Spin-dipole Giant Resonance}
Recently it has been proposed
 to utilize the excitation of the spin-dipole resonance, excited in charge
exchange
 reactions,  to determine
 the neutron skin; in fact the method has been applied to obtain information
on the variation of
 the neutron skin in the Sn isotopes with isotope number \cite{Kras99}.
For the relevant operator, $\sum \sigma_i\tau_i^{\pm} r_i Y_1(r_i), $
the summed $\Delta L=1$ strength is % sensitive to $\Delta R$
  \begin{eqnarray}
& S^{(-)}-S^{(+)}=C(N R^2_n-Z R^2_p).
\label{1}
\end{eqnarray}
Here $S^{(-)}$ and $S^{(+)}$ are the spin-dipole total strengths
in $\beta^{(-)}$ and $\beta^{(+)}$ channels, respectively; $C$ is
the factor depending on the definition of the spin-dipole operator
(in the definition of Ref.~[2] $C=1/4\pi$, we use here $C=1$).
Because $S^{(+)}$ could not be measured experimentally, the
model-dependent
energy-weighted sum rule %($EWSR$)
 was invoked in the analysis to eliminate $S^{(+)}$.
Let put $S^{(+)}=0$ (that seems to be a very good approximation for $^{208}$Pb)
 and ask the question
what experimental accuracy for $S^{(-)}$ is needed to determine
the neutron skin to a given accuracy. With
%With $r_p=\sqrt{\langle r^2\rangle_p}$,
%be the neutron skin thickness.
\begin{eqnarray}
& S^{(-)}=(N-Z) R^2_p+2N R_p\Delta R \label{S-}
\end{eqnarray}
%The first term on rhs is fixed.
 the ratio of the second term on the rhs to
the first one in case of  $^{208}$Pb is
%\begin{eqnarray}
$$ 2N\Delta R /((N-Z)  R_p)\approx 5.7 \Delta R / R_p. $$
%\nonumber
%\end{eqnarray}
%
Therefore, for $R_p=5.5$ fm and $\Delta R=0.2$ fm the second term is only 25\%
of the
first one and one needs 3\% accuracy in $S^{(-)}$ to determine $\Delta R$ with
10\% accuracy.
Because the SDR strength is spread out and probably has a considerable
strength at low-energy
%(not observed  experimentally),
the results for the $\Delta R$ can be only considered as
qualitative  with a relatively large uncertainty (of order 30-50\%).

\subsubsection{Isobaric analogue state}

The dominant contribution to the energy weighted sum rule (EWSR)
 for Fermi excitations by the operator   $T^{(-)}$
comes from the Coulomb mean field \cite{Auer72}
\be
(EWSR)_F= \int U_C(r)n^{(-)}(r)d^3r, \label{1.11}
\ee
The Coulomb mean field $U_C(r)$ resembles very much
that of the uniformly charged sphere, being inside a nucleus a quadratic
function: $U_C(r)=\displaystyle\frac{Ze^2}{2R_c}(3-(r/R_c)^2), \ r\le R_c$. It
turns out that
if one extends such a quadratic dependence also to the outer region  $r>R_c$
(instead of proportionality to $R_c/r$),
it gives numerically just very small deviation in $(EWSR)_F$ (less than 0.5\%,
due to the fact, that
the difference and its first derivative go to 0 at $r=R_c$ and $n^{(-)}(r)$ is
exponentially decreasing
at $r>R_c$). Using such an approximation, one gets:
\begin{eqnarray}
&&(EWSR)_F\approx (N-Z)\Delta_C\left(1-\frac{S^{(-)}}{3(N-Z)R^2_c}\right)
\label{1.22}
\end{eqnarray}
with $\Delta_C=\displaystyle\frac{3Ze^2}{2R_c}$, and $S^{(-)}$
given in Eq.(\ref{S-}).

Since the IAS exhausts almost 100\% of the NEWSR and EWSR, one may
hope to extract $S^{(-)}$ from the IAS energy. However, the term
depending on $S^{(-)}$ contributes only about 20\% to $(EWSR)_F$,
and as a result, the part of $S^{(-)}$ depending on $\Delta R$
contributes only about 4\% to $(EWSR)_F$ (in $^{208}$Pb). %%
>From the experimental side, the IAS energy can be determined with
unprecendently high accuracy, better than
0.1\%. Also, from the experimentally known charge density distribution
 the Coulomb mean field $U_C(r)$ can be
calculated rather accurately, and hence one can determine
 the small difference between Eqs.(\ref{1.22}) and (\ref{1.11}).
But at the level of 1\% accuracy several theoretical effects
discarded in Eq.(\ref{1.11}) come into play (see, e.g.,
\cite{Auer72}) which makes  the reliability of such a method
questionable. On the other hand in a forth coming paper we will
show that for an isotopic chain the excitation of the IAS can be
used as a quantitative tool to obtain the variation of $\Delta R$
with neutron number.

%%%%%%%%%%%%%%%%%%%%%%%%%%%%%%%%%%%%%%%%

\subsection{Anti-protonic atoms}
Recently neutron density distributions were deduced from
anti-protonic atoms \cite{Trzc01}. The basic method determines the
ratio of neutron and proton distributions at large differences by
means of a measurement of the annihilation products which
indicates whether the antiproton was captured on a neutron or a
proton. In the analysis two  assumptions are made. First
 a best fit value for the ratio $R_I$ of the imaginary parts of
the free space $\bar{p}p$ and $\bar{p}n$ scattering lengths equal
to unity is adopted. Second in order to reduce the density ratio
at the annihilation side to a a ratio of rms radii a two-parameter
Fermi distribution is assumed. The model dependence introduced by
this assumptions is difficult to judge. Since a large number of
nuclei have been measured one may argue that the value of $R_I$ is
fixed empirically.
% The value obtained  for  $\Delta R$ for $^{208}$Pb is
%$0.15 \pm 0.02$ fm.
%%%%%%%%%%%%%%%%%%%%%%%%%%%%%%%%%%%%%%%%%
 \subsection{Parity violating electron scattering}
Recently it has been proposed to use the (parity violating) weak
interaction to probe the neutron distribution. This is probably
the least model dependent approach \cite{Horopv}. The weak
potential between electron and a nucleus \be \tilde{V}(r)=
V(r)+\gamma_5A(r), \ee where the axial potential $A(r)=
\frac{G_F}{2^{3/2}} \rho_W(r). $  The weak charge is mainly
determined by neutrons \be \rho_W(r)= ( 1-4\sin^2\theta_W)
\rho_p(r)- \rho_n(r), \ee with $\sin^2\theta_W  \approx 0.23. $ In
a scattering experiment using polarised electrons one can
determine the cross section asymmetry \cite{Horopv} which comes
from the interference between the $A$ and $V$ contributions. Using
the measured neutron form factor at small finite value of $Q^2 $
and the existing information on the charge distribution one can
uniquely extract the neutron skin. Some slight model dependence
comes from the need to assume a certain radial dependence for the
neutron density,  to extract $R_n$ from a finite $Q^2$ form
factor.
%%%%%%%%%%%%%%%%%%%%%%%%%%%%%%%%%%%%%%
\section{Discussion of $\Delta R$ for $^{208}$Pb and some  implications}
\label{Discussion} In  table \ref{summaryDeltaR} we present a
summary of some recent results on $\Delta R$ in $^{208}$Pb. One
sees that (with the exception of  the
 analysis of proton and neutron pickup reactions in
terms of mean field orbitals in \cite{Mairle}) all recent results
are consistent with $\Delta R \sim 0.13\pm 0.03 $ fm.
%The analysis
%of a  new measurement on the excitation of the isovector giant
%dipole resonance will lead to a smaller statistical error \cite{Kras02}.
Therefore  it appears that the data agree with the result of
conventional Skyrme model approach but seem to disagree with the results
of the RMF models considered in \cite{Furn}.
On the basis of the correlation plot between $\Delta R$ and $p_0$
shown in \cite{Furn} on would then conclude that a small value for
$p_0 \sim 2 $MeV/fm$^3$ is preferred over larger values prdedicted in RNF
approaches.

\begin{table} \label{summaryDeltaR}
\begin{center}
\caption{Summary of recent results for $\Delta R$ in $^{208}$Pb}
\begin{tabular}{c|ccc}
method & $\Delta R$ [fm] & error [fm] & ref\\
 \hline giant dipole resonance excitation  & 0.19 &
0.09 & \cite{Kras94,Kras02} \\
neutron/proton pickup & 0.51 &  &      \cite{Mairle} \\
($\vec{p},p^{\prime})$ at 0.5-1.04 GeV   & 0.097 & 0.014 &  \cite{Clark03} \\
nucleon scattering (40-200 MeV) &  0.17 &   & \cite{Karat02} \\
anti-protonic atoms & 0.15 & 0.02 & \cite{Trzc01} \\
%pion &  &  & \\
parity violating electron scattering  & planned & 0.05 &
\cite{Horopv}
\\ \hline
%\caption{Exp results for $\Delta R$}
\end{tabular}

\end{center}
\end{table}
 In several processes of physical interest
knowledge of  $\Delta R$ plays a crucial role and in fact a more
accurate value could lead to more stringent tests:

(i) The pion polarization operator \cite{Kolo} (the s-wave optical
potential) in a heavy nucleus $
\Pi(\omega,\rho_p,\rho_n)=-T^+(\omega)\rho -T^-(\omega) (
\rho_n-\rho_p) $ has mainly an isovector character
 ($T^+(m_\pi) \sim 0).$
 Parameterizing the
densities by Fermi shapes for the case of $^{208}$Pb the main
nuclear model dependence in the analysis comes  from the
uncertainty in  the value of $\Delta R$ multiplying $T^-.$

 (ii) Parity violation in atoms is dominated by $Z-$boson
exchange between the electrons and the neutrons \cite{Poll99,
Derev}. Taking the proton distribution as a reference there is a
small socalled neutron skin (ns) correction to the parity
non-conserving amplitude, $\delta E_{\mbox{pnc}}^{\mbox{ns}},$
for, say, a $6s_{1/2} \to 7s_{1/2}$ transition, which is related
to $\Delta R$ as \cite{Derev} (independent of the electronic
structure) \be \frac {\delta E^{\mbox{ns}}_{\mbox{pnc}}}{
E_{\mbox{pnc}}}= -\frac{3}{7} (\alpha Z)^2 \frac {\Delta R}{R_p}.
\ee In $^{133}$Cs it amounts to a $\delta E/E \approx -(0.1-
0.4)\% $ depending on whether the non-relativistic or relativistic
estimates for $\Delta R $ are used \cite{Poll99}. The
corresponding uncertainty in the weak charge  $Q_W$ is
$-(0.2-0.8)\sigma.$

 (iii) The  pressure in  neutron star matter can be expressed as
in terms of the symmetry energy and its density dependence \cite{Latt}
\be P(\rho,x)= \rho^2 \frac{\partial E(\rho,x)} {\partial \rho} =
\rho^2 [ E'(\rho, 1/2) +S'(\rho) (1-2x)^2 +\ldots ]. \label{press}
\ee By using the beta equilibrium condition in a neutron star, $\mu_e=
\mu_n-\mu_p= -\frac{\partial E(\rho,x)}{\partial x}, $ and the
result for the
 electron chemical potential, $\mu_e=3/4\hbar cx(3\pi^2\rho x)^{1/3}, $ one
finds the
proton fraction near saturation density, $\rho_0,$ to be quite
small,
 $x_0 \sim 0.04.$
Hence the  pressure at saturation density  can be approximated as
\be P(\rho_0) = \rho_0(1-2x_0)(\rho_0S'(\rho_0)(1-2x_0)+ S(\rho_0)
x_0) \sim \rho^2_0 S'(\rho_0). \ee
At higher densities the proton
fraction increases; this increase is more rapid in case of larger
$p_0$ \cite{Horo02}. While for the pressure at higher densities
contributions from other nuclear quantities like compressibility
will play a role in \cite{Latt}
%Using the Tolman-Oppenheimer-Volkov equation
 it was argued that that there is a  correlation of the neutron star
 radius and the pressure % mass $M$
which does not depend on the EoS at the highest densities.
 %($M=1.4 M_{solar}$)
 Numerically the correlation can be expressed in the form of a power
 law,
 $R_M \sim C(\rho,M) (\frac{P(\rho)}{\rm{MeV fm}^{-3}})^{0.25}$
 km,
where $C(\rho=1.5 \rho_0,M=1.4 M_{solar})  \sim 7 . $ This shows
that a determination of a neutron star radius would provide some
constraint on the symmetry properties of nuclear matter.

\section{Conclusion}
In this paper we have discussed the bulk symmetry energy, and
compared various approaches to compute it as a function of
density. Because the tensor interaction plays an important role
the symmetry energy is sensitive to details of the  treatment of
the many-body correlations.  It was shown that  the
self-consistent Green function approach in which more correlations
are included than in lowest order BHF leads to a smaller value of
the symmetry energy. The relatively large values for $p_0$
obtained in the relativistic mean field approach can be associated
with an effective mass effect.

 We showed that the phenomenological almost linear relationship
between symmetry energy and neutron skin in finite nuclei observed
in mean field calculations could be understood in terms the
Landau-Migdal approach.
 Finally we compared several experimental tools of potential
 interest for the determination of the neutron skin.
\\ \\ {\bf Acknowledgement} \\
This work is part of the research program of the ``Stichting voor
Fundamenteel Onderzoek der Materie" (FOM) with financial support
from the ``Nederlandse Organisatie voor Wetenschappelijk
Onderzoek" (NWO). Y. Dewulf acknowledges support from FWO-Vlaanderen.
 \\ The authors would like to thank Prof. M.Urin
for several clarifying remarks.

\end{document}